\documentclass[12pt]{article}
\def\be{\begin{equation}}
\def\ee{\end{equation}}
\def\bea{\begin{eqnarray}}
\def\eea{\end{eqnarray}}

\begin{document}
\begin{flushright}
{ IFT/01-06}\\
{ SINP/TH/00-33}\\
{ TIFR/TH/01-06}\\
hep-ph/0103161
\end{flushright}

\begin{center}
{\Large{\bf Supersymmetric lepton flavour violation 
in a linear collider: the role of charginos}} \\[2cm]

{\large M. Guchait\footnote{guchait@tnp.saha.ernet.in}}\\[0.3 cm]
{\it Saha Institute of Nuclear Physics}\\
{\it Bidhan Nagar, Calcutta - 700064, India.}
\\[0.75cm]
{\large J. Kalinowski\footnote{Jan.Kalinowski@fuw.edu.pl}}\\[0.3 cm]
{\it Instytut Fizyki Teoretycznej, Uniwersytet Warszawski }\\
{\it Ho\.za 69, 00681 Warszawa,  Poland.}
\\[0.75cm]
{\large Probir Roy\footnote{probir@tifr.res.in}}\\[0.3 cm]
{\it Tata Institute of Fundamental Research}\\
{\it Homi Bhabha Road, Mumbai-40005, India.}\\
\end{center}

\bigskip

\begin{abstract}
The occurrence of a significant amount of supersymmetric lepton
flavour violation at laboratory energies, through $\tilde\nu_\mu -
\tilde\nu_\tau$ mixing, has become a realistic possibility in the wake
of the super-Kamiokande atmospheric neutrino result.  This effect can
be observed in an $e^+e^-$ linear collider with the distinct final
state $\tau \mu+ jets + {E\!\!\!/}_T$.  We show that the pair
production of charginos can make an important contribution to this
process and has to be taken into account in addition to that of
sneutrinos or charged sleptons.  Some case studies are presented with
CM energies of 500 and 800 GeV and integrated luminosities of 50, 500
and 1000 ${\rm fb}^{-1}$.
\end{abstract}

\newpage

\vspace{1in}

\section{Introduction.} 
The results of the super-Kamiokande atmospheric neutrino experiment
\cite{atm} provide compelling evidence of lepton flavour violation.
Combined with recent data from reactor antineutrino studies
~\cite{antinu}, these strongly suggest a large near-maximal mixing
$(\theta_{\nu_\mu \nu_\tau} \sim \pi/4)$ and consequent oscillations
between very lightly massive mu and tau neutrinos.  The latest
analysis implies $\Delta m^2 \equiv |m^2_{\nu_\mu} - m^2_{\nu_\tau}|
\sim 3 \times 10^{-3} ~{\rm eV}^2$ and $\sin^2
2\theta_{\nu_\mu\nu_\tau} > 0.88$.  Unless the two neutrinos are
closely degenerate, their masses may be expected to have the same
order of magnitude as their mass difference.  The long-standing
deficit of solar neutrinos \cite{solar} may also be explained by
neutrino oscillations, though the presence of $\nu_e \rightarrow
\nu_\mu$ or $\nu_e \rightarrow \nu_\tau$ oscillations at a high level
is still an open question.

Neutrino oscillations imply the violation of individual lepton flavour
numbers and raise an interesting possibility of observing processes
with a violation of lepton flavour between two charged leptons, such
as $\mu\rightarrow e\gamma$, or $\tau\rightarrow\mu\gamma$
\cite{muegamma,muth}. In the Standard Model these processes are
strongly suppressed due to the GIM mechanism. However, in the
supersymmetric extension of the Standard Model, new mechanisms with
virtual superpartner loops may enhance \cite{muth} these rare decay
processes. Of course, once superpartners are discovered, it will be
possible to probe lepton flavour violation directly in their
production and decay processes \cite{kras}.  For example, it has been
demonstrated that sneutrino or charged slepton pair production at
future $e^+e^-$ (and/or $\mu^+\mu^-$) colliders may provide a more
powerful tool to search for supersymmetric lepton flavour violation
(SLFV) than the said rare decay processes \cite{feng,nojiri}.

In this note we point out that sneutrinos and charged sleptons may not
only be directly pair-produced in $e^+e^-$ collisions, but can also be
decay products of other supersymmetric particles, like charginos and
neutralinos, decaying via cascades. The latter may contribute to the
signal as well as background for SLFV processes.  Therefore, a
detailed account of these is needed in assessing the sensitivity of
future colliders to SLFV.  We find that off-diagonal chargino
pair-production, overlooked earlier, can make a significant
contribution to the SLFV signal already at $\sqrt{s} = 500$ GeV, and
further that the role of neutralinos in decay chains is quite
important and needs to be taken into account. At a CM energy of 800
GeV the diagonal pair production of the heavier chargino may also need
to be included. We provide detailed studies of the SLFV signal at
these two energies for two representative points of the parameter
space of the MSSM. Significance contours are drawn in the parameter
plane with the sneutrino mass difference as one axis and the
sine of twice the sneutrino mixing angle as the other.

\section{Signatures of slepton mixing} 
Within the framework of a seesaw mechanism \cite{seesaw}, it is
reasonable to suppose that masses and mixings in the
$\nu_\mu-\nu_\tau$ system are caused by very heavy right-handed
Majorana neutrinos with masses that are upwards of $10^{10}$ GeV.
Then flavour violating mixings get radiatively induced~\cite{slfv} in
the charged slepton and sneutrino sectors via renormalization group
equations.  In such a scheme a substantial $\nu_\mu - \nu_\tau$ mixing
leads to~\cite{feng,hisano} large $\tilde\mu_L - \tilde\tau_L$ and
$\tilde\nu_\mu - \tilde\nu_\tau$ mixings. In this paper we do not
discuss these theories; we concentrate on the question how well 
SLFV can be probed at future $e^+e^-$ colliders in a model independent
way.

For nearly degenerate sleptons, additional SLFV contributions to rare
decay processes are suppressed as $\Delta m_{\tilde{l}}/m_{\tilde{l}}$
through the superGIM mechanism. As a result, constraints from the yet
unobserved radiative decays $\mu \rightarrow e\gamma$, $\tau
\rightarrow \mu \gamma$, $\tau \rightarrow e\gamma$ are not very
stringent \cite{muegamma}.  On the other hand, in decays of sleptons,
this kind of supersymmetric lepton flavour violation is suppressed
only as $\Delta m_{\tilde{l}}/\Gamma_{\tilde{l}}$ \cite{feng}.  Since
$m_{\tilde{l}}/\Gamma_{\tilde{l}}$ is typically of the order
$10^2$--$10^3$, one may expect spectacular signals \cite{feng,nojiri}
for possible discovery in future $e^+e^-$ or $\mu^+ \mu^-$ collider
experiments.

To be more specific, let us take a pure 2-3 intergeneration mixing
between $\tilde\nu_\mu$ and $\tilde\nu_\tau$, generated by a
near-maximal mixing angle $\theta_{23}$, and let us ignore any mixings
with $\tilde\nu_e$. This means that scalar mass matrices are not
diagonal in the same basis as fermion mass matrices.  For example, the
scalar neutrino mass matrix $m^2_{\tilde\nu}$, restricted to the 2-3
generation subspace, can be written in the fermion mass-diagonal basis
as
\begin{equation}
m^2_{\tilde\nu} = \left(\matrix{\cos\theta_{23} & -\sin\theta_{23} \cr
\sin\theta_{23} & \cos\theta_{23}}\right) 
\left(\matrix{m_{\tilde\nu_2} & 0 \cr 0 &
m_{\tilde\nu_3}}\right) 
\left(\matrix{\cos\theta_{23} & \sin\theta_{23} \cr
-\sin\theta_{23} & \cos\theta_{23}}\right),
\label{one}
\end{equation}
where $m_{\tilde\nu_2}$ and $m_{\tilde\nu_3}$ are the physical masses
of $\tilde\nu_2$ and $\tilde\nu_3$ respectively.  In the following we
take the mixing angle $\theta_{23}$ and $\Delta m_{23} =
|m_{\tilde\nu_2} - m_{\tilde\nu_3}|$ as free, independent
parameters. The same goes for the charged slepton sector, modulo
standard LR mixing, where $\theta_{23}$ and $\Delta m_{23}$ are then
the corresponding parameters for charged sleptons.\footnote{One can
give a parallel discussion for the $e$-$\tau$ mixing case
\cite{nomura}, replacing $\mu$ by $e$ everywhere.} If we work in the
mass eigenstate basis for all fields, the slepton mixing matrix will
appear in interaction vertices of sleptons with leptons and
charginos/neutralinos.  As a result, the SLFV signals can be looked
for in decays of sleptons, for example
\begin{eqnarray}
e^+e^- & \rightarrow &
\tilde{\ell}^-_i\tilde{\ell}^+_i \rightarrow  \tau^+\mu^- \tilde{\chi}^0_1
\tilde{\chi}^0_1, \nonumber \\
e^+e^- & \rightarrow &
\tilde{\nu}_i\tilde{\nu}^c_i  \rightarrow  \tau^+\mu^- \tilde{\chi}^+_1
\tilde{\chi}^-_1 \label{slelfv}
\end{eqnarray}
with $i=2,3$, or in decays  of charginos and/or neutralinos
\begin{eqnarray}
e^+e^- & \rightarrow &
\tilde\chi^+_2\tilde{\chi}^-_1   \rightarrow  \tau^+\mu^- 
 \tilde\chi^+_1\tilde\chi^-_1  \label{charlfv}\\
e^+e^- & \rightarrow &
\tilde\chi^0_2  \tilde{\chi}^0_1 \rightarrow  \tau^+\mu^-
\tilde\chi^0_1\tilde\chi^0_1 \label{neulfv}
\end{eqnarray}
where $\tilde{\chi}^\pm_1 \rightarrow \tilde{\chi}^0_1 f\bar{f}'$, and
$\tilde{\chi}^0_1$ escapes detection.  The signature therefore would
be $\tau^{\pm}\mu^{\mp}+ jets+ {E\!\!\!/}_T$,  $\tau^{\pm}\mu^{\mp}+
\ell + {E\!\!\!/}_T$, or $\tau^{\pm}\mu^{\mp}+ {E\!\!\!/}_T$, 
depending on hadronic or leptonic
$\tilde{\chi}^\pm_1$ decay mode.

Hisano {\it et al.} \cite{nojiri} first proposed and discussed the search
for SLFV signals (through $\tilde\nu_\mu - \tilde\nu_\tau$ mixing) at
an assumed $e^+e^-$ or $\mu^+\mu^-$ linear collider with $\sqrt{s} =
500$ GeV and $\int dt {\cal L} = 50 ~{\rm fb}^{-1}$ considering only
the pair-production of sneutrinos and of charged sleptons, eq.(\ref{slelfv}).  
The final states, generated by SLFV and analyzed by them, were
\begin{equation}
{\rm (A):}\; \tau \mu+ 4j+ {E\!\!\!/}_T, \qquad\quad
{\rm (B):}\; \tau \mu l +2j+{E\!\!\!/}_T, \qquad\quad
{\rm (C):}\; \tau \mu l \bar l +2j+ {E\!\!\!/}_T 
\label{ABC}
\end{equation}
(with $l=e,\mu$ and$j=$jets coming from chargino decays) 
for which the corresponding backgrounds
were small and under control.  They found that only the signal (A) is
viable in an $e^+ e^-$ collider whereas the other two processes would
be difficult to observe due to small rates.  The observability of signal
(B) is better at a $\mu^+ \mu^-$ collider, whereas the signal (C) is
less promising in this collider too.  In their analysis they assumed
$m_{\tilde\chi_1^\pm}=100$ GeV.

However, if the chargino $\tilde\chi^\pm_2$ is not much heavier, as is
the case in a substantial region of the MSSM parameter space, then
off-diagonal chargino pair production $e^+e^- \rightarrow
\tilde\chi^\pm_1 \tilde\chi^\mp_2$ can take place for the linear
collider CM energy $\sqrt{s}=500 ~{\rm GeV}$. The heavier chargino can
decay via the SLFV chain, eq.(\ref{charlfv}), and the chargino
$\tilde\chi^\pm_1 \tilde\chi^\mp_2$ pair production can also lead to
the same final states as in eq.(\ref{ABC}) providing a new source for
the signal in addition to those discussed in Ref.\cite{nojiri}.
Moreover, the production of two charginos in $e^+e^-$ collision has
both the $s$-channel and $t$-channel exchange contributions and hence
expected somewhat larger cross sections at higher collider energies.
Other production processes, like $\tilde\chi^\pm_2 \tilde\chi^\mp_2$,
$\tilde\chi^0_i \tilde\chi^0_j$, may also be open at higher energies
(depending on the mass pattern) and contribute to the same final
states not only via the SLFV mechanisms shown in eqs.(\ref{charlfv})
and (\ref{neulfv}), but through lepton flavour conserving decay chains
as well. The main difference between our study and that of the Hisano
et.al \cite{nojiri} is that realistically we have taken into account
additional contributions to SLFV coming from chargino production.  We
have also considered background from production of supersymmetric
particles which were not considered in Ref. \cite{nojiri}.  Moreover,
since we allow two jets in (A) to overlap, we also consider an
important SM background coming from $\bar{t}tg$ production followed by
semileptonic top decays.  We note that recently possible signals of
the SLFV in chain decays of neutralinos produced at the LHC have been
discussed \cite{lfvatlhc}.

\begin{table}[!t]
\begin{center}
\begin{tabular}{|c||c|c|c||c|c|c|}
\hline
& \multicolumn{3}{c||}{ RR1} & \multicolumn{3}{c|}{RR2}\\
\hline
Particle & Mass & Decay  & BR & Mass & Decay  & BR\\
\hline\hline
$\tilde\chi^+_1$ & 128 &$\tilde\chi^0_1 \ell^+ \nu_\ell$ & 0.15 $\times$ 3 
& 132 & $\tilde{\tau}^+\nu_\tau$ &1.0\\
&& $\tilde\chi^0_1 q \bar{q}'$ & 0.56 &&& \\
\hline
$\tilde\chi^+_2$ & 346 & $\tilde\chi^0_2 W^+ $ & 0.29 
&295 & $\tilde{\chi}^0_2 W^+$ & 0.31\\
&&$\tilde\chi^+_1 Z$ & 0.22 
& &$\tilde\chi^+_1 Z$ & 0.22\\
&&$\tilde\chi^+_1 h$ & 0.14
& &$\tilde\chi^+_1 h$ & 0.13 \\
&&$\tilde{t}_1 \bar{b}$      & 0.14
& &$\tilde{\nu}_\tau \tau^+$& 0.07 \\
&&$\tilde\ell^+ \nu_\ell$ & 0.04 $\times$ 3 
& &$\tilde{\tau}^+_2 \nu_{\tau}$ & 0.08 \\
&&$\ell^+\tilde{\nu}_\ell$ & 0.03 $\times$ 3
& & $\tilde{\chi}^0_1W^+$ & 0.06 \\
\hline
$\tilde\chi^0_1$ & 72 & & & 75 &&\\
\hline
$\tilde\chi^0_2$ & 130 & $\tilde\chi^0_1 \tau^+\tau^-$ & 0.24 
& 133 & $\tilde{\tau}^+\tau^-$ & 0.50\\
&&$\tilde\chi^0_1 e^+ e^-$ &0.20
& & $\tilde{\tau}^-\tau^+$ & 0.50 \\
&&$\tilde\chi^0_1 \mu^+ \mu^-$ &0.20 
& & &\\
&&$\tilde\chi^0_1 \nu_\ell \bar{\nu}_\ell $ &   0.04 $\times$ 3 &&&\\
\hline
$\tilde\chi^0_3$ & 320 & $\tilde\chi^{\pm}_1 W^{\mp}$ & 0.62
& 273 & $\tilde\chi^{\pm}_1 W^{\mp}$ & 0.54 \\
&&  $\tilde\chi^0_2 Z$& 0.20
& & $\tilde\chi^0_2 Z$& 0.15\\
&&  $\tilde\chi^0_1 Z$& 0.14
& & $\tilde\chi^0_1 Z$& 0.28\\
\hline
$\tilde\chi^0_4$ & 348 & $\tilde\chi^{\pm}_1 W^{\mp}$ & 0.52 
& 293 &  $\tilde\chi^{\pm}_1 W^{\mp}$ & 0.52 \\
&& $\tilde\chi^0_2 h^0$ & 0.11
& &  $\tilde\chi^0_2 h^0$ & 0.09\\
&& $\tilde\chi^0_1 h^0$ & 0.07
& &  $\tilde\chi^0_1 h^0$ & 0.06\\
\hline
$\tilde{\ell}^-_L$ & 176 & $\tilde{\chi}^-_1\nu_{\ell}$& 0.53
& 217 & $\tilde{\chi}^-_1\nu_{\ell}$ & 0.52\\
&&  $\tilde\chi^0_2 \ell^-$ & 0.32 
&&  $\tilde\chi^0_2 \ell^-$ & 0.34 \\
&&  $\tilde\chi^0_1 \ell^-$ & 0.15 
&&  $\tilde\chi^0_1 \ell^-$ & 0.14 \\
\hline
$\tilde{\nu}_{\ell}$ & 161 &$\tilde{\chi}^0_1\nu_{\ell}$& 0.48
& 202 & $\tilde{\chi}^+_1\ell^-$ & 0.55\\ 
&&  $\tilde\chi^0_2 \nu_{\ell}$ & 0.12 
&&  $\tilde\chi^0_2 \nu_{\ell}$ & 0.20 \\
&&  $\tilde\chi^0_1 \nu_{\ell}$ & 0.40 
&&  $\tilde\chi^0_1 \nu_{\ell}$ & 0.25 \\
\hline
$\tilde{\tau}^-_1$ & 131 & $\tilde{\chi}^0_1 \tau^-$& 1.00
& 92 &  $\tilde{\chi}^0_1 \tau^-$& 1.00\\
\hline
$\tilde{\tau}^-_2$ & 177 &$\tilde{\chi}^-_1 \nu_\tau $& 0.53
& 209 &  $\tilde{\chi}^-_1 \nu_\tau$& 0.38\\
&&   $\tilde\chi^0_2 \tau^-$& 0.31 
&&   $\tilde\chi^0_2 \tau^-$& 0.28 \\
&&   $\tilde\chi^0_1 \tau^-$& 0.16 
&&   $\tilde\chi^0_1 \tau^-$& 0.28 \\
\hline
$\tilde{\nu}_{\tau}$ & 161 &$\tilde{\chi}^0_1\nu_{\tau}$& 0.48
& 177 & $\tilde{\chi}^+_1\tau^-$ & 0.49\\ 
&&  $\tilde\chi^0_2 \nu_{\tau}$ & 0.12 
&&  $\tilde\chi^0_2 \nu_{\tau}$ & 0.17 \\
&&  $\tilde\chi^0_1 \nu_{\tau}$ & 0.40 
&&  $\tilde\chi^0_1 \nu_{\tau}$ & 0.33 \\
\hline
\end{tabular}
\end{center}
\caption{\small The masses (in GeV) and the branching ratios (only for
significant decay modes) for supersymmetric particles which are
relevant to our study. No slepton mixing is assumed.  
$\ell$ denotes $e$ or $\mu$, and $\tau$ unless the 
entry for $\tau$ is explicitly shown.   
The SUSY parameter points RR1 and RR2 are as
specified  in eq.(\ref{eq:msugra}) \cite{rr}.} 
\end{table}

To illustrate the phenomenology of the SLFV process, we estimate the
signal and background rates for two representative points~\cite{rr} in
the MSSM parameter space given in terms of two mSUGRA scenarios chosen
for detailed case studies at the ECFA/DESY linear collider workshop:
\begin{eqnarray}
&&RR1:\quad m_0=100; 
\quad M_{1/2}=200;\quad A_0=0;\quad \tan\beta=3;\quad
{\rm sgn}(\mu)=+ \nonumber \\
&&RR2:\quad m_0=160; 
\quad M_{1/2}=200;\quad A_0=600;\quad \tan\beta=30;\quad
{\rm sgn}(\mu)=+ 
\label{eq:msugra}
\end{eqnarray}
Here the masses and $A_0$ are in GeV, and standard notation is used.
The masses of corresponding chargino, neutralino and slepton states
are shown in Table 1 along with the relevant branching ratios which are
used in our calculations. The listed branching ratios are for the case
of no slepton mixing; the effect of mixing is calculated below.  The
lightest neutralino $\tilde\chi^0_1$ is nearly a bino while
$\tilde\chi^0_2$ and $\tilde\chi^\pm_1$ are largely winos, thereby
being almost degenerate.  For these MSSM points, the squarks are
heavier than 300 GeV.

A glance at  Table 1 shows that in the case of RR2 ({\it i.e.} for
large $\tan\beta$) the lightest chargino decays only leptonically
(without SLFV it decays into $\chi^0_1\tau\nu_\tau$).  
As a result, the signature of
SLFV in $\tilde\chi^\pm_2\tilde\chi^\mp_1$ production 
would be one muon and three taus plus
missing energy. Such a signal might be extremely difficult to extract from
background with four taus or with two muons and two taus for realistic
tau identification efficiencies. This question calls for full
experimental simulations which are clearly beyond the scope of our
analysis.  Since the process in eq.(\ref{charlfv}) will not contribute to
the final states listed in eq.(\ref{ABC}), it can be neglected, as
done in Ref.\cite{nojiri}.  For the case RR1, however, the lightest
chargino has a large branching ratio for hadronic decays. Therefore
taking into account the process eq.(\ref{charlfv}) in addition to
eq.(\ref{slelfv}) may significantly improve sensitivity of 
an $e^+e^-$ linear collider to
SLFV processes.  As it turns out that processes with one or both
$\tilde{\chi}^\pm_1$ decaying leptonically are overwhelmed by
background, in our analyses we consider only signature (A) in
eq.(\ref{ABC}). Allowing  two quark jets to overlap, in the next
section we discuss the final states with $\tau^\pm\mu^\mp + \ge 3
jets$ for signal and background processes in scenario RR1.

\section{Collider signals}
We perform our study of SLFV in a linear collider, such as the
proposed TESLA~\cite{tesla}, with a CM energy $\sqrt{s} = 500$ and
800 GeV and an integrated luminosity $\int dt {\cal L} = 50-1000
~{\rm fb}^{-1}$.  To study the signal and as well as corresponding
background rates we use simple parton level Monte Carlo simulation,
where each parton is treated as a jet.  The viability of this SLFV
signal is studied both for $\sqrt{s} = 500$ GeV and $\sqrt{s} = 800$
GeV.

\bigskip

\noindent {\bf (a) $\sqrt{s}=500~{\rm GeV}$} case:

\bigskip

In this case, given the mass spectrum of Table 1, the off-diagonal
$\tilde\chi_1^\pm \tilde\chi_2^\mp$ pair is the only possibility for
the SLFV signal in chargino production.  Starting with the $\tilde\chi^\pm_1
\tilde\chi^\mp_2$ state, slepton flavour violation can occur in the
heavier chargino $\tilde\chi_2^\mp$ long cascade decay chain. The
entire decay sequence is shown as follows:
\begin{enumerate}
\item[S1:]
$e^+e^- \rightarrow \tilde{\chi}^\pm_2 \tilde{\chi}^\mp_1 $
\begin{eqnarray}
\tilde\chi^+_2  & \rightarrow & \tau^+ \tilde\nu_{2,3}, \qquad  
\tilde\nu_{2,3} \rightarrow \mu^- \tilde\chi^+_1, \qquad
\tilde\chi^+_1 \rightarrow \tilde\chi^0_1 + q + \bar q', \nonumber \\
\tilde\chi^-_1 & \rightarrow &\tilde \chi^0_1 + q + \bar q'
\label{eq:Asg1}
\end{eqnarray}
\item[S2:]
$e^+e^- \rightarrow \tilde{\chi}^\pm_2 \tilde{\chi}^\mp_1 $
\begin{eqnarray}
\tilde\chi^+_2  & \rightarrow & \mu^+ + \tilde\nu_{2,3}, \qquad
\tilde\nu_{2,3} \rightarrow \tau^- \tilde\chi^+_1, \qquad
\tilde\chi^+_1 \rightarrow \tilde\chi^0_1 + q + \bar q', \nonumber \\
\tilde\chi^-_1 & \rightarrow & \tilde \chi^0_1 + q + \bar q'.
\label{eq:Asg2}
\end{eqnarray}
\end{enumerate}
There is, of course, another sequence with the charges reversed.  
The other process for signal (A), which was discussed
in~\cite{nojiri}, is the following:
\begin{enumerate}
\item[S3:]
$e^+e^- \rightarrow \tilde\nu_i \tilde\nu_i^c$ 
\begin{eqnarray}
\tilde\nu_i  & \rightarrow &  \tilde\chi_1^- \tau^+;\qquad
\tilde\chi_1^- \rightarrow \tilde\chi_1^0 + q +\bar q'; \nonumber \\
\tilde\nu^c_i  & \rightarrow & \tilde\chi_1^+ \mu^-;\qquad
\tilde\chi_1^+ \rightarrow \tilde\chi_1^0  + q + q',\qquad\qquad\qquad
\label{eq:Asg3}
\end{eqnarray}
\end{enumerate}
where $i=2,3$.  Notice that in eqs.(\ref{eq:Asg1},\ref{eq:Asg2}) the
slepton flavour violating decay occurs in two ways leading to the the
same final state so that eventually the signal rate gets doubled.

The cross section corresponding to signal processes S1 and S2 can be
written as
\begin{equation}
\sigma(e^+ e^- \rightarrow
\tau \mu+\ge 3 jets) \simeq \chi_{23}
\sin^2 2\theta_{23} \times \sigma_0
\times \epsilon_{BR} \times \epsilon_{\tau_{id}},\qquad\qquad\;
\label{eq:charcs}
\end{equation}
whereas, for S3, it is given by 
\begin{equation}
\sigma(e^+ e^- \rightarrow
\tau \mu+\ge 3 jets) \simeq \chi_{23}(3-4\chi_{23})
\sin^2 2\theta_{23} \times \sigma_0
\times \epsilon_{BR} \times \epsilon_{\tau_{id}}.
\label{eq:snucs}
\end{equation}
The SLFV effect is taken into account \cite{feng} by the factors
$\sin^2 2\theta_{23}$ and
\begin{equation}
\chi_{23} = \frac{x_{23}^2}{2(1+x_{23}^2)}, \qquad 
x_{23} = \Delta m_{23}/\Gamma,
\end{equation}
where $\Gamma$ is the decay width of the sneutrino, which is 0.42 GeV
for our choice of parameter space and assumed to be independent of
flavour.  The difference between eq.(\ref{eq:charcs}) and
eq.(\ref{eq:snucs}) is due to the correlated slepton pair production
in the process S3.

In the above expressions $\sigma_0$ is the corresponding sparticle
pair-production cross section in $e^+e^-$ collision and
$\epsilon_{BR}$ is the product of relevant branching ratios for the
corresponding decay chains assuming no SLFV. The value of
$\epsilon_{BR}$ are easily obtained by consulting Table 1. For example
for S1 we get $\epsilon_{BR}=0.0075$ where the factor 2 accounting two
possible cases of $\chi^\pm_2\chi^{\mp}_1$ is included. The other
factor, $\epsilon_{\tau_{id}}$, is the $\tau$ lepton selection
efficiency.  In our calculation we have considered decays of the
$\tau$ through its hadronic decay modes to products such as
$\pi\nu_\tau$, $a_1\nu_\tau$ and $\rho\nu_\tau$ with a total branching
ratio of 64\%. We have normalized final decay distributions
appropriately taking the polarization of the $\tau$~\cite{bullock}.
Here the $\tau$ is mostly left handed as it couples to
$\tilde\chi_2^\pm$ through gauge interaction, whereas the right handed
coupling is suppressed by its mass as it couples through higgsino part
of $\tilde\chi_2^\pm$.  Assuming the $\tau$ identification efficiency
0.70~\cite{martyn} for these decay modes, we get for the parameter
$\epsilon_{\tau_{id}}=0.45$ including its branching ratios to hadronic
decay modes.  Note that in full MC simulations one could further
improve the signal to background ratio by imposing a cut on the impact
parameter of the muon since the muon in the background processes comes
in most cases from the decay of a $\tau$ which travels some distance
from the production vertex before it decays~\cite{nojiri}.  However,
we have not used such a cut in the present study.

 A point to be noted is that in the chargino decay processes either
the $\tau~$ (eq.\ref{eq:Asg1}) or the $\mu$ (eq.\ref{eq:Asg2}) is the
leading lepton. The energy distribution of each of these leptons is
flat between a maximum and a minimum value. This feature will be
exploited to suppress possible background.

Since we allow (unlike Ref.\cite{nojiri})  two jets to overlap, 
the most dominant Standard Model background to the 
signal 
$\tau\mu + \ge 3 jets$ comes from 
\begin{equation}
e^+ e^- \rightarrow t \bar t g 
\label{eq:qcdbg}
\end{equation}
production followed by semileptonic decays of two top quarks. 
We have computed this process at tree level 
using MADGRAPH \cite{Stelzer:1994ta} 
with energy and isolation cuts discussed below.
There are also flavour-conserving SUSY processes that contribute 
significantly to the  background. 
In the following we list those processes with
their possible decay chains:
\begin{enumerate}
\item[B1:]
$e^+ e^- \rightarrow \tilde\chi^+_2 \tilde\chi^-_1$
\bea
\tilde\chi^+_2  & \rightarrow & \tau^+ + \tilde\nu_\tau; \qquad
\tilde\nu_\tau \rightarrow \tau^-(\rightarrow \mu^-) \tilde\chi^+_1; \qquad
\tilde\chi^+_1 \rightarrow \tilde\chi^0_1 + q + \bar q'; \nonumber \\
\tilde\chi^-_1 & \rightarrow & \tilde \chi^0_1 + q + \bar q'.
\label{eq:Abg1}
\eea
\item[B2:]
$e^+ e^- \rightarrow \tilde\chi^+_2 \tilde\chi^-_1$
\bea
\tilde\chi^+_2  & \rightarrow &  \tau^+ (\rightarrow \mu^+)+ \tilde\nu_\tau;
\qquad \tilde\nu_\tau \rightarrow \tau^- \tilde\chi^+_1; \qquad
\tilde\chi^+_1 \rightarrow \tilde\chi^0_1 + q + \bar q'; \nonumber \\
\tilde\chi^-_1 & \rightarrow & \tilde \chi^0_1 + q + \bar q'.
\label{eq:Abg2}
\eea
\item[B3:]
$e^+ e^- \rightarrow \tilde\nu_i \tilde\nu^c_i$
\bea
\tilde\nu_i  & \rightarrow &  \tilde\chi_1^- \tau^+;\qquad
\tilde\chi_1^- \rightarrow \tilde\chi_1^0 + q +\bar q'; \nonumber \\
\tilde\nu^c_i  & \rightarrow & \tilde\chi_1^+ \tau^-
(\rightarrow \mu^-);\qquad
\tilde\chi_1^+  \rightarrow  \tilde\chi_1^0  + q + \bar q'.\qquad\qquad\qquad\qquad
\label{eq:Abg3}
\eea
\item[B4:]
$e^+ e^- \rightarrow \tilde\tau_2^+ \tilde\tau_2^-$
\bea
\tilde\tau^+_2 & \rightarrow & \tau^+ \tilde\chi_2^0;\qquad
\tilde\chi_2^0 \rightarrow \tilde\chi_1^0 \tau^+(\rightarrow jets)
\tau^-(\rightarrow \mu^-); \qquad\qquad\nonumber \\
\tilde\tau^-_2 & \rightarrow & \nu_\tau \tilde\chi_1^-;\qquad
\tilde\chi_1^- \rightarrow \tilde\chi_1^0 + q + \bar q'.
\label{eq:Abg4}
\eea
\end{enumerate}
In eq.(\ref{eq:Abg4}) the $\tilde\tau_2$ is the heavier physical state
after the mixing between $\tilde\tau_L$ and $\tilde\tau_R$. Notice
that in all background cases B1--B4 the $\mu$ comes from $\tau$ decay after
the $\tau\tau X$ events are produced.

\bigskip

\noindent {\bf (b)$\sqrt{s}=800~{\rm GeV}$} case:

\bigskip

Because of the higher energy, many new sparticle production channels
containing heavier states of charginos and neutralinos open now,
contributing both to signal and background processes.  Cross sections
of these are typically $\simeq \cal O$(fb).  Note that the
$\tilde\nu_i \tilde\nu^c_i$ pair-production will be suppressed at
higher $\sqrt{s}$ since it is an $s$-channel mediated process. As far
as our signal is concerned, a new source is the diagonal heavier
chargino($\tilde\chi_2^\pm \tilde\chi_2^\mp$) pair-production. One of
the $\tilde\chi_2^\pm$ will decay through the flavour violating mode
as shown in eqs.(\ref{eq:Asg1},\ref{eq:Asg2}) while the other
$\tilde\chi_2^\mp$ will decay sequentially to 2 jets and ${E\!\!\!/}_T$
as shown below:
\begin{enumerate}
\item[S4:]
$e^+e^- \rightarrow \tilde{\chi}^\pm_2 \tilde{\chi}^\mp_2 $
\bea
\tilde\chi^+_2  & \rightarrow & \tau^+ \tilde\nu_{2,3}, \qquad  
\tilde\nu_{2,3} \rightarrow \mu^- \tilde\chi^+_1, \qquad
\tilde\chi^+_1 \rightarrow \tilde\chi^0_1 + q + \bar q', \nonumber \\
\tilde\chi_2^- &\rightarrow & \tilde\chi_1^- Z 
\rightarrow \tilde\chi_1^0 q \bar q'\, \nu \bar\nu, \;\;{\mbox or}\;\; 
\tilde\chi_2^-  \rightarrow  \tilde\chi_2^0 W^- \rightarrow 
\tilde\chi_1^0 \nu \bar\nu\, 
q \bar q'
\label{eq:Asg4}
\eea 
\item[S5:]
$e^+e^- \rightarrow \tilde{\chi}^\pm_2 \tilde{\chi}^\mp_2 $
\bea
\tilde\chi^+_2  & \rightarrow & \mu^+ + \tilde\nu_{2,3}, \qquad
\tilde\nu_{2,3} \rightarrow \tau^- \tilde\chi^+_1, \qquad
\tilde\chi^+_1 \rightarrow \tilde\chi^0_1 + q + \bar q', \nonumber \\
\tilde\chi_2^- &\rightarrow & \tilde\chi_1^- Z 
\rightarrow \tilde\chi_1^0 q \bar q'\, \nu \bar\nu, \;\;{\mbox or}\;\; 
\tilde\chi_2^-  \rightarrow  \tilde\chi_2^0 W^- \rightarrow 
\tilde\chi_1^0 \nu \bar\nu\, 
q \bar q'
\label{eq:Asg5}
\eea 
\end{enumerate}

In the case of background processes, there are also new sources which
cannot be neglected.  We find four additional processes  
leading to the
same final states:
\begin{enumerate}
\item[B5:]
$e^+e^- \rightarrow \tilde\chi_2^+ \tilde\chi_2^-$
\bea
\tilde\chi_2^- &\rightarrow& \tilde\chi_1^- Z
, \;  \tilde\chi_1^- h^0 
\rightarrow
\tilde\chi_1^0 \tau \nu_\tau q \bar q, \nonumber \\
\tilde\chi_2^+ &\rightarrow &\tilde\chi_1^+ Z 
, \;  \tilde\chi_1^+ h^0
\rightarrow
\tilde\chi_1^0 \mu \nu_\mu q \bar q 
\eea
\item[B6:]
$e^+e^- \rightarrow \tilde\chi^0_3 \tilde\chi^0_2$ 
\bea
\tilde\chi_3^0 & \rightarrow & \tilde\chi_1^+ W \rightarrow
\tilde\chi_1^0 q \bar q' q \bar q', \nonumber \\
\tilde\chi_2^0 & \rightarrow & \tilde\chi_1^0 \tau^+ \tau^-(\to \mu).
\label{b6}
\eea 
\item[B7:]
$e^+e^- \rightarrow \chi_{2}^0 \chi_4^0$
\bea
\tilde\chi_2^0 &\rightarrow& \chi_1^0 \tau^+\tau^-(\rightarrow \mu^-), 
\nonumber \\
\tilde\chi_4^0 &\rightarrow& 
\chi^{\pm}_1W^{\mp}\rightarrow \chi^0_1 q\bar{q}'\, 
q\bar{q}' \label{b7}
\eea
\item[B8:]
$e^+e^- \rightarrow \chi_{3}^0 \chi_4^0$
\bea
\tilde\chi_3^0 &\rightarrow& \chi_1^{\pm} W^{\mp} \rightarrow \chi^0_1 \mu\nu 
\, q\bar{q}',
\nonumber \\
\tilde\chi_4^0 &\rightarrow& 
\chi^{\pm}_1W^{\mp}\rightarrow \chi^0_1 \tau\nu \,
q\bar{q}'  \label{b8}
\eea

\end{enumerate}

\section{Expected signal and background rates}
We have used the following selection cuts for our events
using simple parton level simulation:
\begin{enumerate}
\item The $\tau$ and $\mu$ are selected with the restriction
  $|\cos\theta_\mu| < $ 0.99 and $|\cos\theta_\tau|<$ 0.96. This cut
  has been applied to avoid leptons which are very close to the beam
  direction ~\cite{martyn}. We have put a selection cut on $E_\tau > $
  2 GeV. 
\item The selection cut on the missing energy is applied as
  $|\cos{\vec\theta}_{miss}| <$ 0.90 in order to avoid missing energy
  which is along the beam pipe~\cite{martyn}.
\item We select leptons to be isolated if the visible energy around
  the cone $\Delta R=0.4$ is less than the maximum (10\% of $E_\ell$,
  1 GeV).  Here $\Delta R = \sqrt{\Delta\phi(\ell,j)^2 +
    \Delta\eta(\ell,j)^2}$, with $\Delta\phi(\ell,j)$ and
  $\Delta\eta(\ell,j)$ being the differences of azimuth and
  pseudorapidity respectively, between any lepton and one of the jets.
\item As mentioned before, we have treated partons as jets without
  including showering and fragmentation effects.  We consider two jets
  as isolated if they pass the selection cuts $\Delta \tilde R =
  \sqrt{\Delta \phi (j,j)^2 + \Delta \eta (j,j)^2} > 0.6$.  Here
  $\Delta\phi(j,j)$ and $\Delta\eta(j,j)$ are the differences of
  azimuth and pseudorapidity respectively between any two jets.
  We accept those jets which are not too close to the beam pipe,
  {\it i.e.}  with $|\cos\theta_{jet}| <$ 0.95, and 
  have energy $E_j>0.05\, (\sqrt{s}-2m_t)$. The last cut originates 
  from the necessity to regulate the   
  IR singularity of the QCD background process, eq.(\ref{eq:qcdbg}).
\item The final muon energy cut is applied to reduce the
  backgrounds of eqs.(\ref{eq:Abg1}--\ref{eq:Abg4}) and 
  eqs.(\ref{b6}--\ref{b8}).  For these
  background processes, the $\mu$ always comes from $\tau$ decay.
  Therefore, this $\mu$ is expected to have less energy than the $\mu$
  of the signal process.  For example, in the case of the signal process S1, 
  the maximum and minimum energies of the
  $\tau$ and the $\mu$ are given by:
\begin{eqnarray} 
E^{(max,min)}_\tau &=&
  E_{\tau}^{rest}(1 \pm \beta_{\tilde\chi}) \gamma_{\tilde\chi}\nonumber \\
  E^{(max,min)}_\mu &=& E_{\mu}^{rest}(1 \pm \beta_{\tilde\nu})
\gamma_{\tilde\nu}
\label{eq:Emu}
\end{eqnarray}
where 
\begin{equation} 
E_{\tau}^{rest} = \frac{m^2_{\tilde\chi^+_2}
- m^2_{\tilde\nu}} {2 m_{\tilde\chi^+_2}};\quad E_{\mu}^{rest} =
\frac{m^2_{\tilde{\nu}} - m^2_{\tilde\chi^+_1}} {2 m_{\tilde\nu}} 
\end{equation}
are the $\tau$ and $\mu$ energies in the rest frames of
$\tilde{\chi}^+_2$ and $\tilde{\nu}$, respectively, and $\beta$'s and
$\gamma$'s are the respective CM boost parameters.  For the case of
eq.(\ref{eq:Asg2}), $\tau$ and $\mu$ have to be interchanged in the
above expressions.  These maximum and minimum energies depend on
massive particles involved in the initial and final states of two body
decay subprocesses.  Therefore, putting a cut on the energy of $\mu$,
such as $E_\mu > 25$ GeV, reduces substantially these backgrounds.
\end{enumerate}

Note that the dominant SM background, eq.(\ref{eq:qcdbg}), contains two
b-quark jets coming from top decays in contrast to the signal
processes which contain jets initiated by light quarks. Therefore,
further background suppression could be achieved by imposing a veto to
the events containing tagged b jets.  However, we have not used this criteria
in our analysis as it requires a detailed MC simulation including detector
effects which is beyond the scope of the present paper.

Applying the selection cuts, as described by 1-5 above, we have
estimated the signal and background rates for the scenario RR1.

\bigskip

\noindent {\bf (a) $\sqrt{s}=500~{\rm GeV}$} case:

\bigskip

In Table 2 we show the expected cross sections for the signal and
background processes (all cross sections are in ab).  The numbers in
the first row correspond to the raw production
cross sections including branching ratio and $\tau$ identification
factors, whereas the numbers in consecutive rows show the effect of  
kinematic cuts described above. 
Therefore, by using eq.(\ref{eq:charcs}) the total signal
cross sections for S1 and S2 turn out to be:

\begin{table}[!t]
\begin{center}
\begin{tabular}{|c||c|c|c||c|c|c|c|}
\hline
  & S1   & S2  & S3   & B1 & B2 & B3  & B4   \\ \hline
0 & 118  & 118 & 1080 & 20 & 20 & 183 & 173  \\
1 & 118  & 105 & 1035 & 18 & 18 & 176 & 147  \\
2 & 98   & 98  & 945  & 17 & 17 & 161 & 134  \\
3 & 84   & 84  & 765  & 14 & 14 & 122 & 121  \\
4 & 57   & 71  & 661  & 9  & 12 & 107 & 99    \\
5 & 57   & 71  & 508  &  2 & 8  & 27  & 38   \\
\hline
\end{tabular}
\end{center}
\caption{\small The signal and background cross sections (in ab) for
$\sqrt{s}=500$ GeV and the reference point RR1 after each set of cuts
as discussed in the text. The branching ratios and
$\epsilon_{\tau_{id}}$ factors are included.}
\end{table}

\begin{equation}
\sigma(e^+ e^- \rightarrow
\tau \mu+\ge 3 jets) \simeq \chi_{23}
\sin^2 2\theta_{23} \times 0.128\;\; {\rm fb}, 
\label{eq:s1}
\end{equation}
whereas for signal S3, using eq.(\ref{eq:snucs}), it is  
\begin{equation}
\sigma(e^+ e^- \rightarrow
\tau \mu+\ge 3 jets) \simeq \chi_{23}(3-4\chi_{23})
\sin^2 2\theta_{23} \times 0.508\;\; {\rm fb}.
\label{eq:s2}
\end{equation}

In comparison, the total cross section for the background is 0.282~fb
which includes 0.075~fb from SUSY processes listed in Table 2 and
0.207~fb from the QCD process eq.(\ref{eq:qcdbg}) after all cuts. Using
Poisson distributions, the significance is $\sigma_d =
\frac{N}{\sqrt{N+B}}$ where N and B is the number of signal and
background events respectively for a given luminosity. In Fig.1 we
have shown the region (to the right of the curve) in the $\Delta
m_{23} - \sin2\theta_{23}$ plane that can be explored or ruled out at
a 3$\sigma$ level by the linear collider of energy 500 GeV for the
given integrated luminosity. We have drawn the contours for three
luminosity options, namely 50 fb$^{-1}$, 500 fb$^{-1}$ and 1000
fb$^{-1}$, whereas the dashed line shows the reach of the process 
$\tilde{\nu}_i\tilde{\nu}^c_i$ alone, eq.(\ref{eq:Asg3}), using our cuts and 
assuming luminosity of  500 fb$^{-1}$. Comparing the dashed line with line 
B we see that  that the chargino contribution, 
eq.(\ref{eq:Asg1},\ref{eq:Asg2})
increases the sensitivity range to $\sin^2\theta_{23}$ by 10-20\% while 
the sensitivity to $\Delta m_{{23}}$ does not change appreciably. 
\bigskip

\noindent {\bf (b) $\sqrt{s}=800~{\rm GeV}$} case:

\bigskip

Here we have generated events using our parton level simulation and
applied the same selection criteria, as discussed earlier.  Our
estimated signal and background rates are given in Table 3.  
We find that 
the total signal rate in this case is 0.231~fb where the contribution 
from both off diagonal and as well
as diagonal chargino production is 0.123~fb and from sneutrino
production it is 0.108~fb. On the other hand, the background cross
section is 0.315~fb, which includes 0.145~fb from SUSY processes, as given 
by Table 3, and 0.17~fb from 
the QCD process eq.(\ref{eq:qcdbg}).  Using eqs.(\ref{eq:charcs})
and (\ref{eq:snucs}), we can isolate the region which can be probed or
ruled out at a linear collider CM energy of 800 GeV.  This is shown
for two luminosity options in Fig. 2 in the same plane as in Fig.1; 
the luminosity of 50 $fb^{-1}$  is not sufficient to exclude 
any region in the figure for this CM energy.
\begin{table}[!t]
\begin{center}
\begin{tabular}{|c||c|c|c|c|c||c|c|c|c|c|c|c|c|}
\hline
  &S1  &S2  &S3  &S4 &S5 &B1 &B2 &B3  &B4  &B5  &B6 &B7  & B8 \\
\hline
0 &108 &108 &720 &53 &53 &18 &18 &115 &138 &131 &98 &56 &147 \\
1 &101 &101 &720 &51 &51 &17 &17 &107 &125 &123 &89 &54 &138 \\
2 &91  &88  &630 &39 &45 &16 &16 &100 &117 &112 &80 &52 &125 \\
3 &78  &78  &360 &38 &38 &13 &13 &61  &95  &102 &80 &51 &115 \\
4 &25  &31  &130 &38 &37 &5  &5  &22  &36  &102 &15 &10  &53  \\
5 &25  &31  &108 &30 &37 &1  &4  &7   &18  &70  &4  &4   &32  \\
\hline
\end{tabular}
\end{center}
\caption{\small The signal and background cross sections (in ab) for
$\sqrt{s}=800$ GeV and the reference point RR1 after each set of cuts
as discussed in the text. The branching ratios and
$\epsilon_{\tau_{id}}$ factors are included.}
\end{table}

\section{Conclusions}

In this paper we have discussed the detection of SLFV at a future
$e^+e^-$ linear collider. We assumed the dominant $\nu_\mu -\nu_\tau$
mixing as suggested by super-Kamiokande atmospheric results. In the
MSSM, the SLFV in left-handed sleptons is induced radiatively, leading
to second and third generation slepton mixing, and to a final state
with $\tau\mu+\ge 3 jets+{E\!\!\!/}_T$ in $e^+e^-$ collisions as a
most interesting signal.  The physics potential of exploring SLFV at
an $e^+e^-$ LC has been analyzed for two reference points in the
supersymmetry parameter space: RR1 with low $\tan\beta=3$, and RR2
with high $\tan\beta=30$, and for two CM energies $\sqrt{s}=500$ and
800 GeV.  The novel feature of our analysis is the inclusion of
diagonal and non-diagonal chargino and neutralino 
pair production processes to the
SLFV signal and background.  The calculations have been performed at
the parton-level Monte Carlo simulations including realistic
experimental cuts. The SM background from the QCD process
$e^+e^-\rightarrow t \bar{t}g$ has been included as well.

For the low $\tan\beta=3$ scenario (RR1) we find that, though the
sneutrino pair production process is the dominant one, the chargino
pair production processes are not negligible contributing more to the
signal than to the background. As a result, the sensitivity of the 
linear collider 
to the SLFV in the parameter space of $\Delta m_{23}-\sin2\theta_{23}$
plane is increased.  We also find that operating at
$\sqrt{s}=500$ GeV, where only non-diagonal $\tilde\chi_1^\pm
\tilde\chi_2^\mp$ chargino pairs can be produced, is the most
optimal for the RR1 scenario. 
Increasing the CM energy to 800 GeV reduces the sensitivity mainly
due to the decrease of sneutrino cross section not compensated by
opening $\tilde\chi_2^\pm \tilde\chi_2^\mp$.
The latter also contributes significantly to the background (B5)
at $\sqrt{s}= 800$ GeV, thereby diluting the SLFV signal at that 
energy. Thus $\sqrt{s} = 800$ GeV does not seem to be a viable CM energy 
option for studying SLFV in the RR1 scenario.

For the high $\tan\beta=30$ scenario (RR2), only the
$\mu+3\tau+{E\!\!\!/}_T$ final states might signal SLFV processes.
However, only detailed Monte Carlo simulations including all
experimental aspects may tell us if such final states can
realistically be reconstructed.

\vskip0.25cm\noindent \underbar{\sl Acknowledgements} \\ We thank
M. Drees, H. Martyn, M.M. Nojiri, X. Tata and P. Zerwas for helpful
discussions.  MG is also thankful to V.Ravindran for fruitful
discussion.  The authors acknowledge the hospitality of the DESY
theory group. MG was supported by Alexander Von Humboldt Foundation 
during the initial phase of this work. The work of JK has partially 
been supported by the KBN
Grant No. 2 P03B 060 18. Work supported in part by the European
Commission 5-th framework contract HPRN-CT-2000-00149.

\def\baselinestretch{1.0}
\vfill\clearpage\thispagestyle{empty}
\noindent
{Figure 1: The significance contours (for the SUSY point RR1 mentioned in
 the text) in $\Delta m_{23} -\sin2\theta_{23}$ plane for
 $\sqrt{s}=$500 GeV and for different luminosity options, contours A,
 B and C being for 50 fb$^{-1}$, 500 fb$^{-1}$ and 1000 fb$^{-1}$,
 respectively. The dashed line is for only $\tilde\nu \tilde\nu^c$
 contribution with luminosity 500 fb$^{-1}$.  The upper-right side of
 these contours can be explored or ruled out at the 3$\sigma$ level. }

\vspace{12cm}
\begin{center}  
\includegraphics{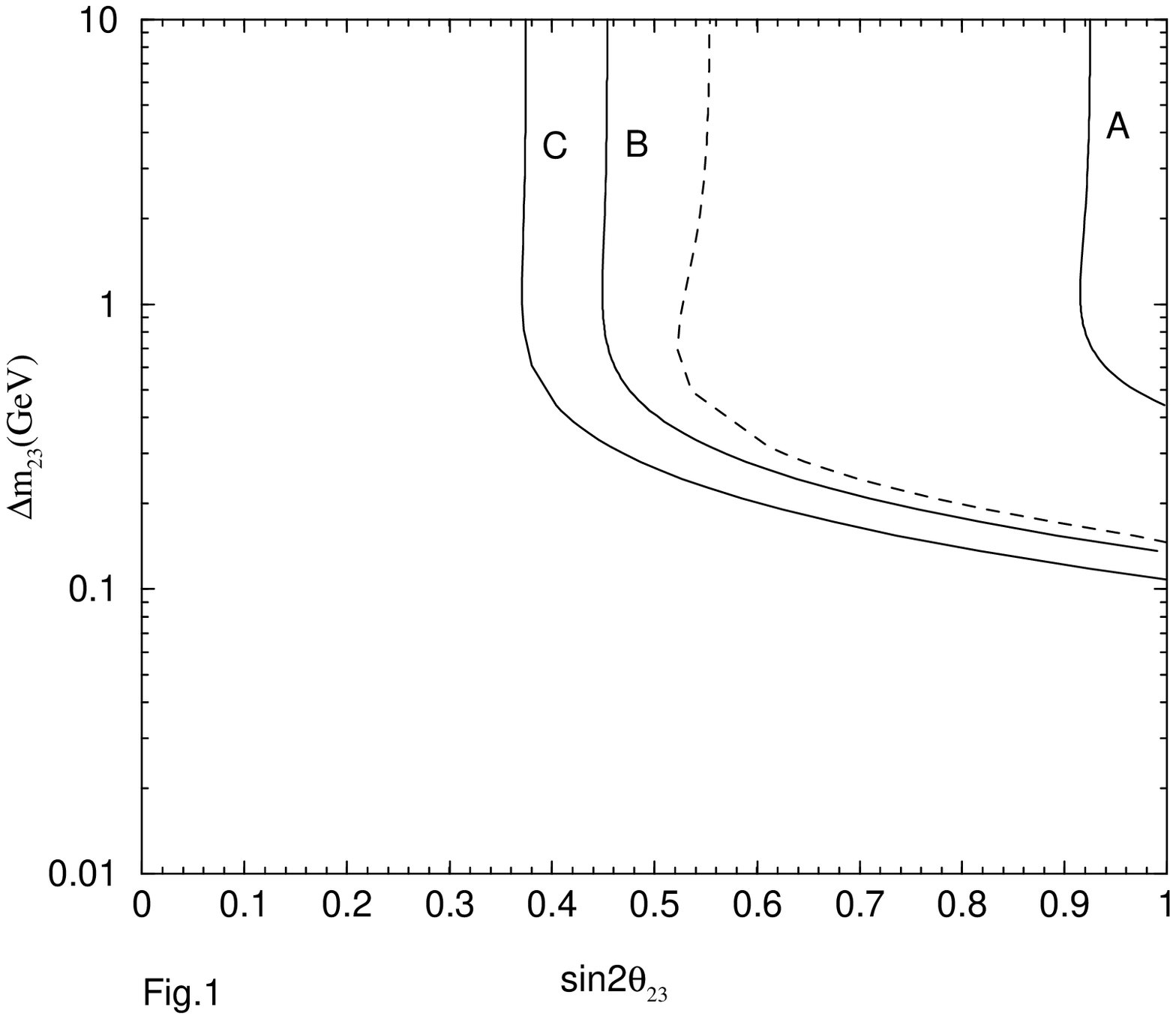}
\end{center}

\vfill\clearpage\thispagestyle{empty}
\noindent
{Figure 2: The same as in Fig.1 for  $\sqrt{s}=$800 GeV and two luminosities  
 500 fb$^{-1}$ and 1000 fb$^{-1}$}  

\vspace{12cm}
\begin{center}
\includegraphics{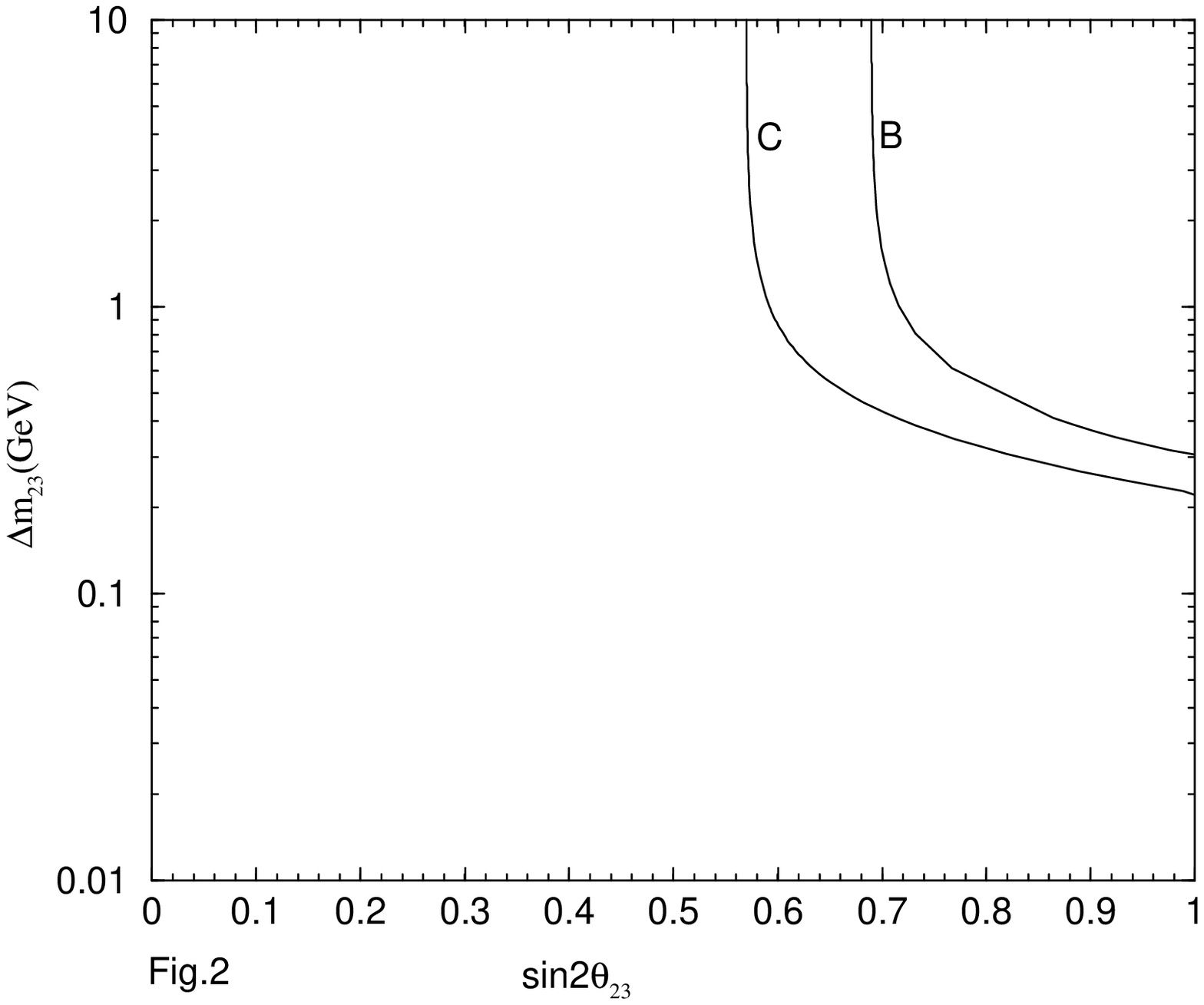}
\end{center}

\end{document}